%% file: UCNSpinEchoLetter.tex
\documentclass[aps,prl,superscriptaddress,showkeys]{revtex4}

\usepackage{amsmath}
\usepackage{amssymb}
\usepackage[dvipsnames,usenames]{color}
\usepackage[pdftex]{graphicx,epsfig}

\usepackage{lmodern}
\usepackage[T1]{fontenc}
\usepackage[english]{babel}

\usepackage[caption=false]{subfig}
\usepackage{epstopdf}
\usepackage{bm}
\usepackage{wasysym}
\usepackage{textcomp}
\usepackage{gensymb}
\usepackage{ulem}
\usepackage{xspace}
\usepackage[tight,nice]{units}
\usepackage[breaklinks=true,hidelinks]{hyperref}

\input{./CommandsAndShortcuts.tex}

\newcommand{\del}[1]{}

\begin{document}

\title{Observation of gravitationally induced vertical striation of polarized ultracold neutrons by spin-echo spectroscopy}

\author{S.~Afach}
\affiliation{Paul Scherrer Institut, CH-5232 Villigen PSI,
Switzerland}
\affiliation{ETH Z\"{u}rich, Institute for Particle Physics, CH-8093 Z\"{u}rich, Switzerland}
\affiliation{Hans Berger Department of Neurology, Jena University Hospital, D-07747 Jena, Germany}

\author{N.J.~Ayres}
\affiliation{Department of Physics and Astronomy, University of Sussex, Falmer, Brighton BN1 9QH, UK}

\author{G.~Ban}
\affiliation{LPC Caen, ENSICAEN, Universit\'{e} de Caen, CNRS/IN2P3, Caen, France}
\author{G.~Bison}
\affiliation{Paul Scherrer Institut, CH-5232 Villigen PSI,
Switzerland}

\author{K.~Bodek}
\affiliation{Marian Smoluchowski Institute of Physics, Jagiellonian University, 30-059 Cracow, Poland}
%
\author{Z.~Chowdhuri}
\affiliation{Paul Scherrer Institut, CH-5232 Villigen PSI,
Switzerland}


\author{M.~Daum}
\affiliation{Paul Scherrer Institut, CH-5232 Villigen PSI,
Switzerland}

\author{M.~Fertl}
\altaffiliation[Present address:  ]{University of Washington, Seattle, United States of America}

\author{B.~Franke}
\altaffiliation[Present address:  ]{Max-Planck-Institute of Quantum Optics, Garching, Germany.}
\affiliation{Paul Scherrer Institut, CH-5232 Villigen PSI,
Switzerland}
\affiliation{ETH Z\"{u}rich, Institute for Particle Physics, CH-8093 Z\"{u}rich, Switzerland}

\author{W.C.~Griffith}
\affiliation{Department of Physics and Astronomy, University of Sussex, Falmer, Brighton BN1 9QH, UK}

\author{Z.D.~Gruji\'c}
\affiliation{Physics Department, University of Fribourg, CH-1700 Fribourg, Switzerland}

\author{P.G.~Harris}
\affiliation{Department of Physics and Astronomy, University of Sussex, Falmer, Brighton BN1 9QH, UK}

\author{W.~Heil}
\affiliation{Institut f\"{u}r Physik, Johannes-Gutenberg-Universit\"{a}t, D-55128 Mainz, Germany}

\author{V.~H\'{e}laine}
\altaffiliation[Present address: ]{LPSC, Universit\'{e} Grenoble Alpes, CNRS/IN2P3, Grenoble, France}
\affiliation{Paul Scherrer Institut, CH-5232 Villigen PSI,
Switzerland}
\affiliation{LPC Caen, ENSICAEN, Universit\'{e} de Caen, CNRS/IN2P3, Caen, France}

\author{M.~Kasprzak}
\affiliation{Physics Department, University of Fribourg, CH-1700 Fribourg, Switzerland}

\author{Y.~Kermaidic}
\affiliation{LPSC, Universit\'{e} Grenoble Alpes, CNRS/IN2P3, Grenoble, France}

\author{K.~Kirch}
\affiliation{Paul Scherrer Institut, CH-5232 Villigen PSI,
Switzerland}
\affiliation{ETH Z\"{u}rich, Institute for Particle Physics, CH-8093 Z\"{u}rich, Switzerland}

\author{P.~Knowles}
\altaffiliation{Present address: LogrusData, Rilkeplatz 8, Vienna, Austria}
\affiliation{Physics Department, University of Fribourg, CH-1700 Fribourg, Switzerland}

\author{H.-C.~Koch}
\affiliation{Physics Department, University of Fribourg, CH-1700 Fribourg, Switzerland}
\affiliation{Institut f\"{u}r Physik, Johannes-Gutenberg-Universit\"{a}t, D-55128 Mainz, Germany}

\author{S.~Komposch}
\affiliation{Paul Scherrer Institut, CH-5232 Villigen PSI,
Switzerland}
\affiliation{ETH Z\"{u}rich, Institute for Particle Physics, CH-8093 Z\"{u}rich, Switzerland}

\author{A.~Kozela}
\affiliation{Henryk Niedwodniczanski Institute for Nuclear Physics, Cracow, Poland}

\author{J.~Krempel}
\affiliation{ETH Z\"{u}rich, Institute for Particle Physics, CH-8093 Z\"{u}rich, Switzerland}

\author{B.~Lauss}
\affiliation{Paul Scherrer Institut, CH-5232 Villigen PSI,
Switzerland}

\author{T.~Lefort}
\author{Y.~Lemi\`{e}re}
\affiliation{LPC Caen, ENSICAEN, Universit\'{e} de Caen, CNRS/IN2P3, Caen, France}

\author{A.~Mtchedlishvili}
\affiliation{Paul Scherrer Institut, CH-5232 Villigen PSI,
Switzerland}

\author{M.~Musgrave}
\affiliation{Department of Physics and Astronomy, University of Sussex, Falmer, Brighton BN1 9QH, UK}

\author{O.~Naviliat-Cuncic} \altaffiliation[Present address:  ]{Michigan State University, East-Lansing, USA.}
\affiliation{LPC Caen, ENSICAEN, Universit\'{e} de Caen, CNRS/IN2P3, Caen, France}

\author{J.M.~Pendlebury}
\affiliation{Department of Physics and Astronomy, University of Sussex, Falmer, Brighton BN1 9QH, UK}

\author{F.M.~Piegsa}
\affiliation{ETH Z\"{u}rich, Institute for Particle Physics, CH-8093 Z\"{u}rich, Switzerland}

\author{G.~Pignol}
\affiliation{LPSC, Universit\'{e} Grenoble Alpes, CNRS/IN2P3, Grenoble, France}

\author{C.~Plonka-Spehr}
\affiliation{Institut f\"{u}r Kernchemie, Johannes-Gutenberg-Universit\"{a}t, Mainz, Germany}

\author{P.N.~Prashanth}
\affiliation{Instituut voor Kern- en Stralingsfysica, University of Leuven, B-3001 Leuven, Belgium}

\author{G.~Qu\'{e}m\'{e}ner}
\affiliation{LPC Caen, ENSICAEN, Universit\'{e} de Caen, CNRS/IN2P3, Caen, France}

\author{M.~Rawlik}
\affiliation{ETH Z\"{u}rich, Institute for Particle Physics, CH-8093 Z\"{u}rich, Switzerland}

\author{D.~Rebreyend}
\affiliation{LPSC, Universit\'{e} Grenoble Alpes, CNRS/IN2P3, Grenoble, France}

\author{D.~Ries}
\affiliation{Paul Scherrer Institut, CH-5232 Villigen PSI,
Switzerland}
\affiliation{ETH Z\"{u}rich, Institute for Particle Physics, CH-8093 Z\"{u}rich, Switzerland}

\author{S.~Roccia}
\affiliation{CSNSM, Universit\'{e} Paris Sud, CNRS/IN2P3, Orsay, France}

\author{D.~Rozpedzik}
\affiliation{Marian Smoluchowski Institute of Physics, Jagiellonian University, 30-059 Cracow, Poland}

\author{P.~Schmidt-Wellenburg}
\email[Corresponding autor: ]{philipp.schmidt-wellenburg@psi.ch} \affiliation{Paul Scherrer Institut, CH-5232 Villigen PSI,
Switzerland}

\author{N.~Severijns}
\affiliation{Instituut voor Kern- en Stralingsfysica, University of Leuven, B-3001 Leuven, Belgium}
\author{J.A.~Thorne}
\affiliation{Department of Physics and Astronomy, University of Sussex, Falmer, Brighton BN1 9QH, UK}
%
\author{A.~Weis}
\affiliation{Physics Department, University of Fribourg, CH-1700 Fribourg, Switzerland}

\author{E.~Wursten}
\affiliation{Instituut voor Kern- en Stralingsfysica, University of Leuven, B-3001 Leuven, Belgium}

\author{G.~Wyszynski}
\author{J.~Zejma}
\affiliation{Marian Smoluchowski Institute of Physics, Jagiellonian University, 30-059 Cracow, Poland}

\author{J.~Zenner}
\affiliation{Paul Scherrer Institut, CH-5232 Villigen PSI,
Switzerland}
\affiliation{ETH Z\"{u}rich, Institute for Particle Physics, CH-8093 Z\"{u}rich, Switzerland}
\affiliation{Institut f\"{u}r Kernchemie, Johannes-Gutenberg-Universit\"{a}t, Mainz, Germany}

\author{G. Zsigmond}
\affiliation{Paul Scherrer Institut, CH-5232 Villigen PSI,
Switzerland}

\begin{abstract}
We describe a spin-echo method for ultracold neutrons (UCNs) confined in a precession chamber and exposed to a $|B_0|=\unit[1]{\muT}$ magnetic field. We have demonstrated that the analysis of UCN spin-echo resonance signals in combination with knowledge of the ambient magnetic field provides an excellent method by which to reconstruct the energy spectrum of a confined ensemble of neutrons. The method takes advantage of the relative dephasing of spins arising from a gravitationally induced striation of stored UCN of different energies, and also permits an improved determination of the vertical magnetic-field gradient with an exceptional accuracy of \unit[1.1]{pT/cm}. This novel combination of a well-known nuclear resonance method and gravitationally induced vertical striation is unique in the realm of nuclear and particle physics and should prove to be invaluable for the assessment of systematic effects in precision experiments such as searches for an electric dipole moment of the neutron or the measurement of the neutron lifetime.
\end{abstract}

\maketitle

Spin-echo resonances, first observed by Hahn in 1950\,\cite{Hahn1950}, have proven to be a very powerful tool in nuclear magnetic resonance~(NMR) experiments for identifying different sources of depolarization in spin-polarized samples. In particular, the $\pi/2$ -- $\pi$ pulse sequence  proposed by Carr and Purcell\,\cite{Carr1954} is widely applied to distinguish the intrinsic spin-lattice or spin-spin coherence time $T_2$ from the global coherence time $T_2^{\ast}$ dominated by dephasing of precessing spins due to spatially varying magnetic fields. Another established application of the spin-echo technique is in neutron scattering\,\cite{Mezei1972} where it is used to resolve correlation times below \unit[1]{\mus}.
Essentially, the spin-echo technique can be applied in solid and soft matter samples where the intrinsic component is long compared to dephasing, which can even be the case for $T_2\ll\unit[1]{s}$. Two similar components of depolarization\,\cite{Harris2014} exist in experiments using UCNs\,\cite{Baker2006,Ban2007,Serebrov2008,Lamoreaux2009,Altarev2009,Altarev2010,Serebrov2010} where coherence times $T_2^{\ast}$ of several hundred seconds have been achieved. 
The defining characteristic of UCNs is that they are reflected from appropriate material surfaces at all angles of incidence, even at room temperature, conserving kinetic energy as they do so. They can therefore be confined 
for observation times that are commensurate with neutron $\beta$-decay: inelastic scattering with phonons from the storage-vessel walls contributes $\Gamma \approx \unit[10^{-3}]{s^{-1}}$ to their loss rate. This peculiarity also makes them ideal for neutron lifetime measurements. A measurement of the evolution of the energy spectrum in these UCN storage experiments can give insight into energy dependent systematic effects and could help to understand the disagreement between beam and storage experiments\,\cite{Olive2014}.

Ultracold neutrons have kinetic energies $E_\text{kin}$ of the same order of magnitude as their gravitational potential energy $E_\text{pot}$ above the lower confining surface. For any given UCN  $E=E_\text{kin}+E_\text{pot}$ is constant during storage, although for the ensemble the average energy decreases over time because faster neutrons have a higher loss rate. These energy-dependent losses result in a time dependence of both depolarization terms, making $T_2^{\ast}$ a function of the free-precession time. 
%
%
 For neutrons $mg\approx\unit[1.03]{neV/cm}$, so, e.g., UCN with kinetic energy of \unit[200]{neV} can rise about \unit[200]{cm} in the gravitational field. 
Therefore, it is convenient (and common) to refer to the energy $\epsilon$ of UCN in terms of the maximum height (in cm) attainable within Earth's gravitational field, and we do so here during our analysis, although we revert to $E$ (in neV) for our final results.

Two channels of field-induced depolarization of UCNs exist: 
the known intrinsic depolarization due to inhomogeneous magnetic fields
\cite{Cates1988PRA37,McGregor1990,Pendlebury2004,Schmid2008,Golub2011,Pignol2012a}, and an energy-dependent relative dephasing for different energies in a vertical magnetic-field gradient.
The latter shows an interplay between gravitationally defined spatial distributions of the neutrons and a vertical magnetic-field gradient $\partial B_z/\partial z$ and has only recently been studied theoretically\,\cite{Knecht2009,Harris2014,Afach2015PRD}. It is the subject of this investigation which uses a new form of spin-echo spectroscopy that also circumvents effects provoked by the time-dependent softening of the UCN spectrum. Prior to this, the spin-echo resonance method was used in a study of Berry's phase with UCNs to cancel any dynamic phase shift\,\cite{Filipp2009} with a coherence time of approximately \unit[850]{ms}.

We have measured UCN spin-echo signals using an apparatus dedicated to the neutron electric dipole moment~(nEDM) experiment\,\cite{Baker2011} with a vertical magnetic field of $B_0 \approx\unit[(0,0,-1)]{\muT}$ and depolarization times of more than \unit[450]{s}.
These measurements, together with first-order knowledge of the magnetic-field gradients, were used to determine the energy spectrum of the UCN after storage as well as a precise value for a common offset to the vertical magnetic-field gradients. The technique has the potential to analyze and correct for the so-called geometric-phase effect\,\cite{Pendlebury2004,Afach2015ELJD}, which is currently the most important systematic effect in experiments searching for an nEDM\@.\\

The equilibrium density distribution of UCNs as a function of height $h$ above the lower confinement surface in a bottle with diffusely scattering walls is $\rho(h)= \rho(0)\cdot\sqrt{ 1- h/\epsilon}$\,\cite{Pendlebury1994}.
This leads to a center-of-mass offset of $H/2\!-\!\left\langle h(\epsilon)\!\right\rangle$ relative to the center plane of a cylindrical precession chamber of radius $r$ and height $H$, where

\begin{equation}
		\left\langle h(\epsilon) \right\rangle=\left\{ 
				\begin{array}{r@{\quad\text{for}~}l}
						\frac{2}{5}\epsilon & \epsilon<H \\
						\frac{\epsilon\left(\frac{2}{5}-\eta + \frac{3}{5}\eta^{5/3} \right)  }{1-\eta} & \epsilon\geq H 							
				\end{array}\right.
\label{eq:Height}
\end{equation}

\noindent is the time-averaged height of a neutron\,\cite{Harris2014} with $\eta\!=\!(1\!-\!H/\epsilon)^{3/2}$. 

In the presence of a vertical magnetic-field gradient, neutrons with an energy $\epsilon$ will precess with an average Larmor frequency of

\begin{equation}
		\omega(\epsilon) =\gamma_\text{n}
		\left( \pm\left\langle B_0\right\rangle-\Part{B_z}{z}\left(H/2\!-\!  \left\langle h(\epsilon)\right\rangle \right)
		\right),
\label{eqn:UCNSEPrecessionFrequency}
\end{equation}

\noindent  where $\pm\left \langle B_0 \right \rangle$ is the volume-averaged magnetic field magnitude in the precession chamber for $B_0$ up\,$(+)$ or down\,$(-)$, and  $\gamma_\text{n} =\unit[-2\pi\!\cdot\!29.1646943(69)]{MHz/T}$\,\cite{Greene1979} is the gyromagnetic ratio of the neutron. 
This induces an energy-dependent relative phase $\Phi(\epsilon)=\omega(\epsilon)\!\cdot\!T$, where $T$ is the free spin-precession~(FSP) time. It leads to a relative dephasing which, when averaged over the energy spectrum, can be interpreted as a gravitationally enhanced depolarization. This relative dephasing can be studied by applying a $\pi/2$ -- $\pi$ -- $\pi/2$ spin-echo sequence, varying the time $t_1$ at which the $\pi$-pulse of duration $t_{\pi}$ is applied while keeping constant the total duration $T+t_{\pi}$ between the end of the first and start of the second $\pi/2$ pulse.
This second $\pi/2$ pulse, at the end of the sequence, is required to transfer information about the spin-precession phase onto the measurable longitudinal spin component. 
For convenience, we analyze the system in a rotating frame precessing with $\omega_\text{0}=\gamma_\text{n}\!\cdot\!\left\langle B_0 \right\rangle$ and $B_0\!>\!0$
inside the storage volume, and we only consider a linear gradient $\partial B_z/\partial z$ along the primary $B_0$ field axis $z$. Therefore, in the rotating frame, the spins of UCNs with energy $E$ precess with average frequency (to first order in $\partial B_z/\partial z$)

\begin{equation}
	\omega_\text{r}(\epsilon) = -\gamma_\text{n}\Part{B_z}{z}\cdot \left( H/2-  \left\langle h(\epsilon)\right\rangle \right).
\label{eqn:frequencyRot}
\end{equation}

\begin{figure}%
\centering
	\includegraphics[width=0.98\columnwidth]{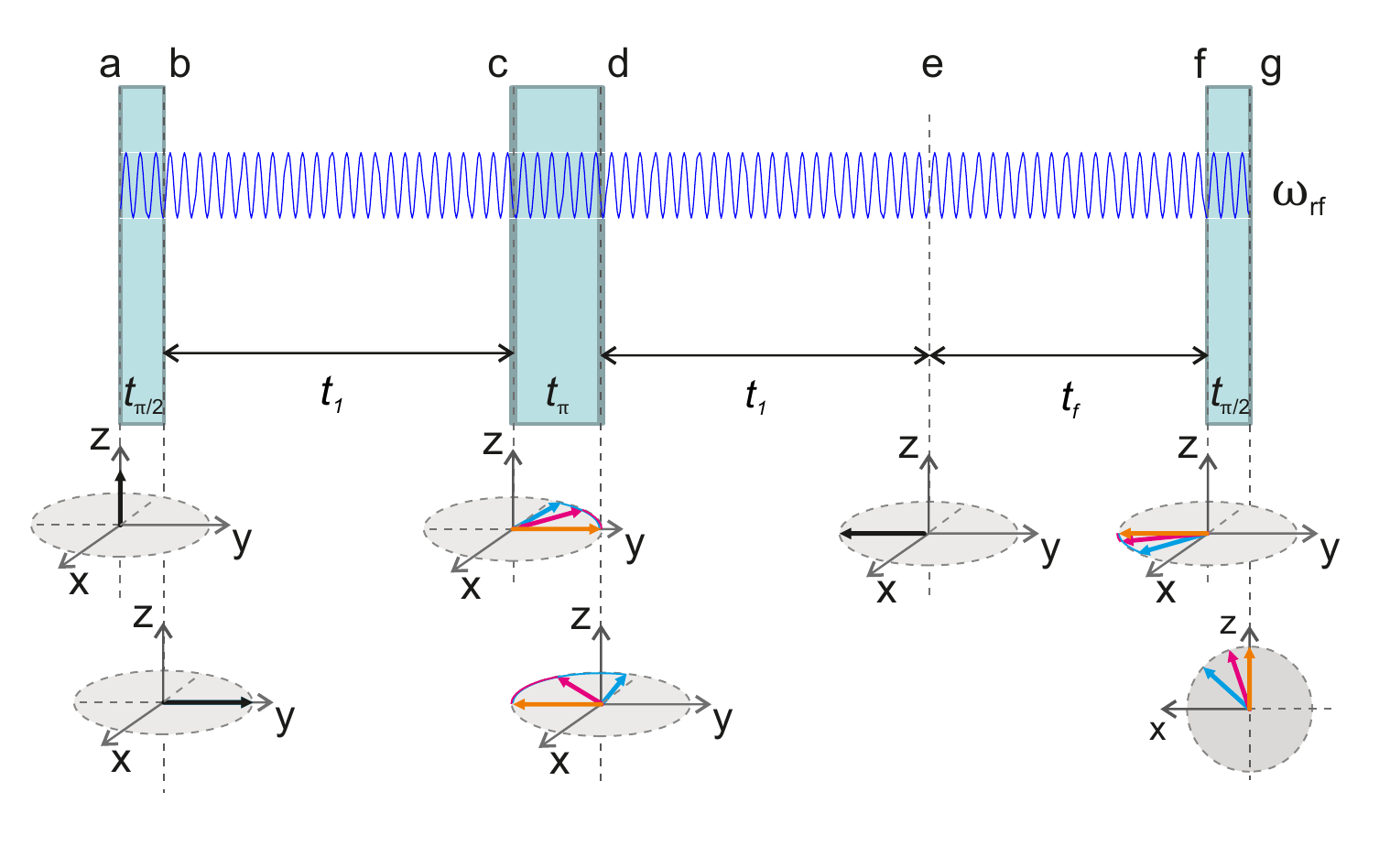}%
	\caption[Illustration of the UCNSE method]
	{Illustration of a UCN spin-echo measurement of duration $T=2t_1\!+\!t_\text{f}$ in the frame rotating at frequency $\omega_\text{rf}=\omega_0$ for a negative gradient $\partial B_z/\partial z$: ({\it a}\,) a polarized ensemble of UCNs is loaded into the cell, 
	({\it a-b}\,) an initial $\pi/2$-pulse tips all spins (black arrow) into the equatorial plane, ({\it b-c}\,) the UCN spins precess with $\omega_\text{r}(E)$ for $t_1$, and fan out. Low-energy UCNs (blue) see a larger field for a negative gradient $\partial B_z/\partial z$ than higher-energy UCNs (magenta), whereas spins that precess at $\omega_0$ 
	(i.e.\ with no center-of-mass offset) are stationary and are oriented along $y$. A $\pi$-pulse ({\it c-d}\,) then flips the spins around the $x$-axis, after which ({\it d-e}\,) the UCNs continue to precess in the same direction, eventually refocusing at $2t_1$ (black arrow).  As they continue to precess beyond this time they fan out again ({\it e-f}\,) until finally ({\it f-g}\,) a second $\pi/2$	pulse is applied. All pulses used for spin manipulation have the same field strength $B_{x}$ applied along the same axis ($x$). A single oscillator is used, with the output being gated on only during the blue time windows; hence, all of the pulses are phase coherent with each other. All energy-dependent effects during the pulses cancel, since a full rotation about $2\pi$ is made and only dephasing during FSP has to be taken into account.}%
\label{fig:UCNSEMethod}%
\end{figure}

\noindent Figure\,\ref{fig:UCNSEMethod} illustrates the evolution of the spins in the rotating frame.
The resonance condition under which the spins of all different energies refocus is established when $T=2t_1$. At this time we measure the highest polarization, which is essentially the polarization $\langle\alpha(T,\epsilon)\rangle_{\epsilon}$ without the gravitationally induced dephasing. If we change the start time $t_1$ of the $\pi$-pulse to values shorter than $T/2$, the resonance condition will have been met before we apply the second $\pi/2$-pulse. As the spins of the different energy classes then proceed to fan out again, the final polarization decreases. Similarly, if the $\pi$-pulse is applied at times larger than $T/2$ the refocusing is not yet complete by the time the second $\pi/2$-pulse is applied.  

For a particular UCN energy $\epsilon$ and fixed $T$ the observed polarization after the second $\pi/2$-pulse can be written as

\begin{equation}
		P(\epsilon,t_1) = \alpha(T,\epsilon)\cdot \cos\left(\omega_\text{r}(\epsilon)\left(T-2t_1\right) \right).
\label{eq:Polarisation1}
\end{equation}

\noindent By integrating over the UCN energy spectrum $p(\epsilon,T)$ at the time of polarization analysis $T$ the observed polarization is then

\begin{equation}
		P(t_1) = \int\alpha(T,\epsilon) \cdot \cos\left[\omega_\text{r}(\epsilon)\left(T-2t_1\right)\right]p(\epsilon,T)\diff \epsilon,
	\label{eq:Polarization2}
\end{equation}

\noindent where $p(\epsilon,T)$ is normalized to 1 for a fixed $T$.

\begin{widetext}
\begin{figure}[floatfix]%
\centering
	\subfloat[]{
			\includegraphics[width=0.48\textwidth]{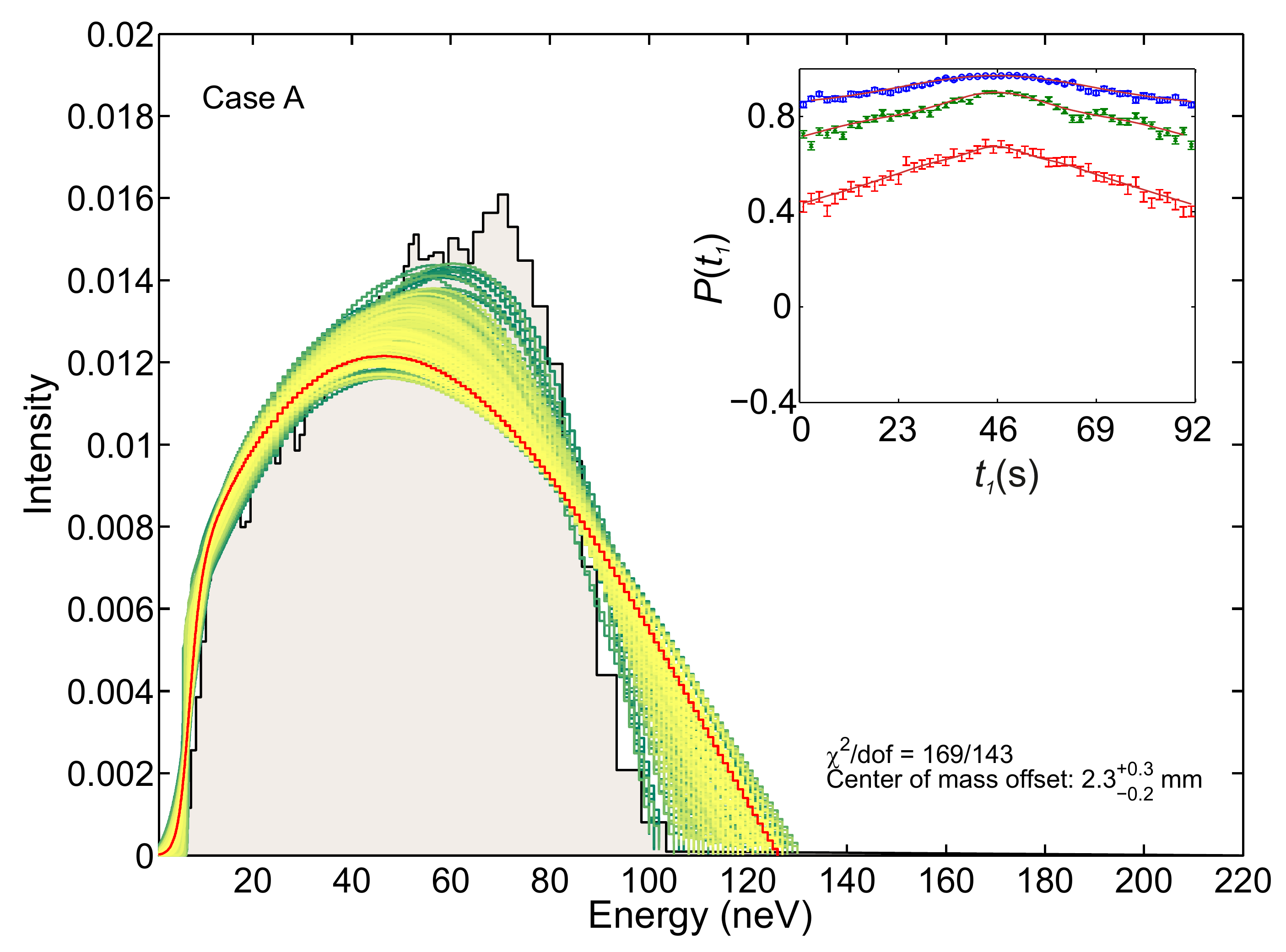}}%
			\hfill
	\subfloat[]{
			\includegraphics[width=0.48\textwidth]{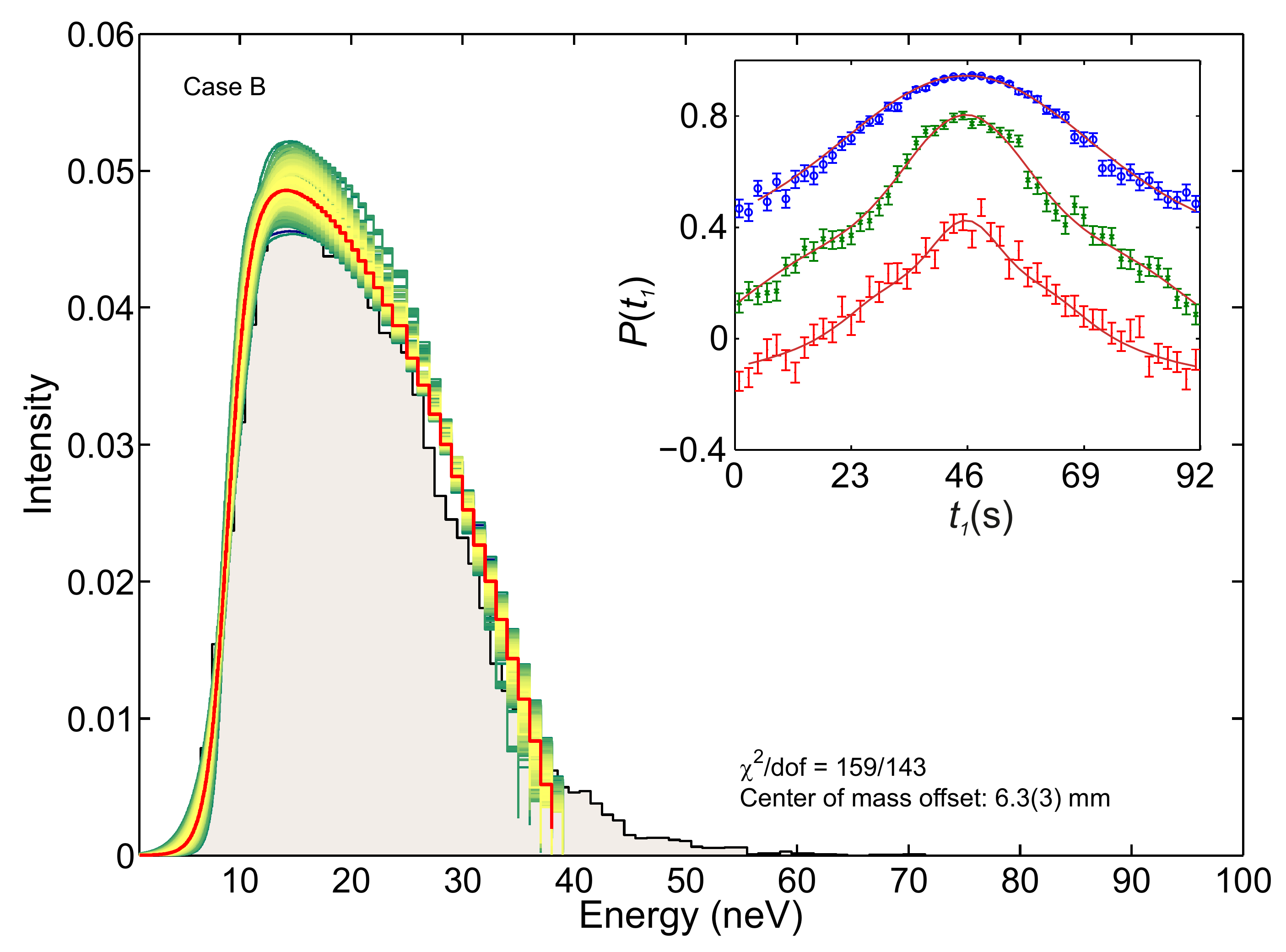}}
			
	\caption{Simulated spin-echo signals and extracted spectra. The inserts show the simulated spin-echo signals using MCUCN\@\cite{Bodek2011}. On each plot three resonances are shown, using the gray input spectra with three different vertical field gradients: blue \unit[100]{pT/cm}; green  \unit[200]{pT/cm}; and red  \unit[400]{pT/cm}. The points with their uncertainties come from the simulation. The red lines on the main plots are the best simultaneous fits to the simulated spectra. The colors (from bright yellow to dark green) represent the $\chi^2$ value for each line; dark green corresponds to a 68.3\% C.L\@ while bright yellow lines have $\chi^2$ values close to the minimum. The original input spectra are represented by the gray areas, which give center-of-mass offsets of \unit[1.93]{mm} (Case A) and \unit[6.46]{mm} (Case B).} %
	\label{fig:SimSpec}%
\end{figure}
\end{widetext}

We have used the MCUCN package\,\cite{Bodek2011} with a detailed model of the nEDM experiment at Paul Scherrer Institute~(PSI)\,\cite{Baker2011} ($H=\unit[12]{cm}$, $r=\unit[23.5]{cm}$) and simulated
the measurement with a $\pi/2$ -- $\pi$ -- $\pi/2$ spin-flip sequence (tip -- flip -- tip) for a fully polarized initial population of UCNs\@. This simulation was carried out for two different UCN energy spectra $p(\epsilon)$, parametrized as in equation\,\eqref{eq:spectrum} below, each with three different constant vertical gradients of $\partial B_z/\partial z = 100$, $200$, and  $\unit[400]{pT/cm}$ for a free precession time of $T=\unit[92]{s}$. The resulting simulated spin-echo resonances are shown in the inserts of Fig.\,\ref{fig:SimSpec} which can be described by equation\,\eqref{eq:Polarization2}. In particular we used
$\alpha(T,\epsilon) = \alpha_0\exp\left(-T\cdot\Gamma_2(\epsilon)\right)$,
\noindent with a transverse relaxation rate as derived in\,\cite{Afach2015PRD} of

\begin{equation}
	 \Gamma_2(\epsilon) = a \frac{\gamma_\text{n}^2}{v\left(\epsilon\right)}
											\left[\frac{8r^3}{9\pi}
											\left(\left|\Part{B_z}{x}\right|^2+\left|\Part{B_z}{y}\right|^2\right)+
											\frac{\mathcal{H}^3(\epsilon)}{16}\left|\Part{B_z}{z}\right|											\right],
\label{eq:Gamma2}
\end{equation}

\noindent where $v(\epsilon)$, and $\mathcal{H}(\epsilon) = \epsilon~\forall~\epsilon\leq H$ and $\mathcal{H}(\epsilon) = H~\forall~\epsilon>H$ are the energy-dependent mean velocity, and the effective height of the UCN as introduced in\,\cite{Pendlebury1994} for a cylindrical cell; $a$ is a free fit parameter.
We found it useful to model the UCN spectrum for a fixed precession time $T$ as 

\begin{equation}
		p(\epsilon) = \frac{A\cdot \epsilon^{1/2}}{1+e^{(E_0-\epsilon)/b^2}}\cdot \left(\frac{1}{1+e^{(\epsilon-(E_0+E_1))/c^2}}\!-\!\frac{1}{1+e^{(E_2-E_0-E_1)/c^2}}\right),
 \label{eq:spectrum}
\end{equation}

 \noindent where $E_0$, $b$ and $E_0+E_1$, $c$ are the points of inflection and slopes of the rising and falling energy edges respectively, while $E_0+E_1+E_2$ is the high-energy cut-off. $A$ is chosen such that $\int^{E_2}_0{p(T,\epsilon)\diff \epsilon}=1$. This energy spectrum is based on a very general distribution $n(\epsilon)\text{d}\epsilon \propto \epsilon^{1/2}\text{d}\epsilon$ from the low-energy tail of a Maxwell-Boltzmann distribution, allowing for low- and high-energy cut-offs. A simultaneous nonlinear least-squares fit (algorithm: `trust-region-reflective') to the signals of all three gradients yielded $a$, $b$, $c$, $E_0$, $E_1$, and $E_2$ for each of the two spectra. The upper and lower confidence intervals were estimated by carrying out random scans of the parameter space around the best value with $\chi^2<\chi^2_\text{limit}$ ($\sim 5\!\times\!10^5$ samples; yellow to dark green curves in Fig.\,\ref{fig:SimSpec}): 

\begin{equation}
	\chi^2_\text{limit}=\chi^2_\text{min}\left(1+\frac{p}{n-p}F_\alpha(n,n-p)\right),
\label{eq:ConfindenceLevel}
\end{equation}

\noindent where $F_\alpha$ is the inverse cumulative function of the statistical $F$-distribution for an $\alpha = \unit[68.3]{\%}$ confidence level with $p=6$ free parameters and $(n-p)=143$ degrees of freedom\,\cite{Rogers1975}. For the numerical integration of equation\,\eqref{eq:Polarization2} we used energy bins of equal height: $\Delta h(\epsilon) = \unit[0.02]{mm}$, which is sufficiently small that any systematic error from the finite energy bins would be less than the statistical resolution of each data point. Figure\,\ref{fig:SimSpec} compares spectra extracted from the fit to the input spectra of the simulation. We note that the results become less reliable at higher energies; this is to be expected, since once the neutrons populate the bottle more or less uniformly it becomes increasingly difficult to discriminate between them.

\begin{figure}%
\centering
	\includegraphics[width=0.98\columnwidth]{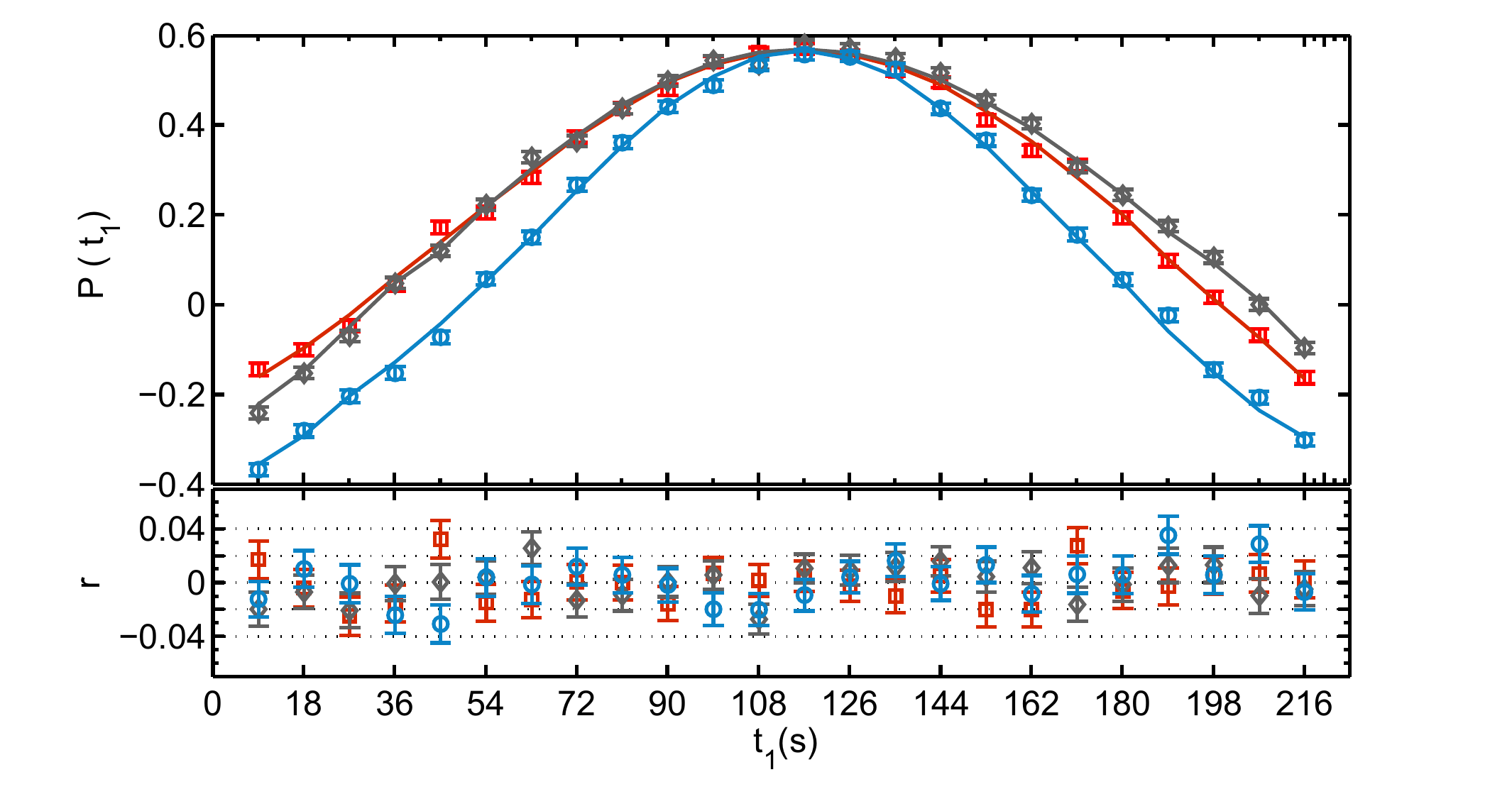}
	\caption[UCN spin-echo measurements with different gradients]
{Measured UCN spin-echo signal for three nominal magnetic field gradients. Red squares: $\partial B_z/\partial z=\unit[-18]{pT/cm}$; black diamonds: \unit[10]{pT/cm}; blue circles:  \unit[38]{pT/cm}. The lines represent the best fit to the data of equation\,\eqref{eq:Polarization2}, using \eqref{eq:spectrum} and \eqref{eq:phase2ndOrder}. The residuals are $r=P(t_1)_\text{fit}-P(t_1)_\text{exp}$.
}%
\label{fig:UCNSE_exp}%
\end{figure}

We measured three spin-echo profiles (see Fig.\,\ref{fig:UCNSE_exp}) using the apparatus described in Ref.\,\cite{Baker2014}, improved with an array of 16 cesium-vapor magnetometers (CsM)\,\cite{Knowles2009} and a simultaneous spin-analyzing system\,\cite{Afach2015}. The magnetic-field measurement with the CsM array permitted us to calculate the magnetic-field gradients using a polynomial field decomposition up to second order, as described in Ref.\,\cite{Afach2014}.
Data were taken for a constant FSP duration of $T=\unit[216]{s}$, with three different trim-coil current settings for $B_0<0$\,(down), thereby applying vertical gradients of $\partial B_z/\partial z = -18$, $10$, and \unit[38]{pT/cm}. The uncertainties of the gradient measurements are dominated by systematics
due to individual sensor offsets, and amount to \unit[12-14]{pT/cm} (see\,\cite{Afach2014,Afach2014b} for a determination of these gradient errors). However, under our conditions, such individual but constant offsets lead to a constant offset $G_z$ of the true $\partial B_z/\partial z$ that can be determined to within \unit[1-2]{pT/cm} using the treatment described below.

The measured initial polarization $\alpha_0 = 0.861(2)$, $0.865(2)$, $0.861(2)$ was prepared by passing UCNs from the UCN source at the PSI\,\cite{Lauss2013} through a \unit[5]{T} solenoid magnet. 
After a filling time of \unit[24]{s}, the UCN shutter in the ground electrode was closed. For each setting, one fixed radio frequency $\omega_\text{rf} = 30.2075, 30.2031$, and \unit[30.2013]{Hz} was used. The \magHg co-magnetometer measured the mean magnetic field $B_0$ via  $\omega_\text{Hg} = \gamma_\text{Hg}\!\cdot\!\langle B_0\rangle$\,\cite{Green1998}.

\begin{figure}%
	\includegraphics[width=0.98\columnwidth]{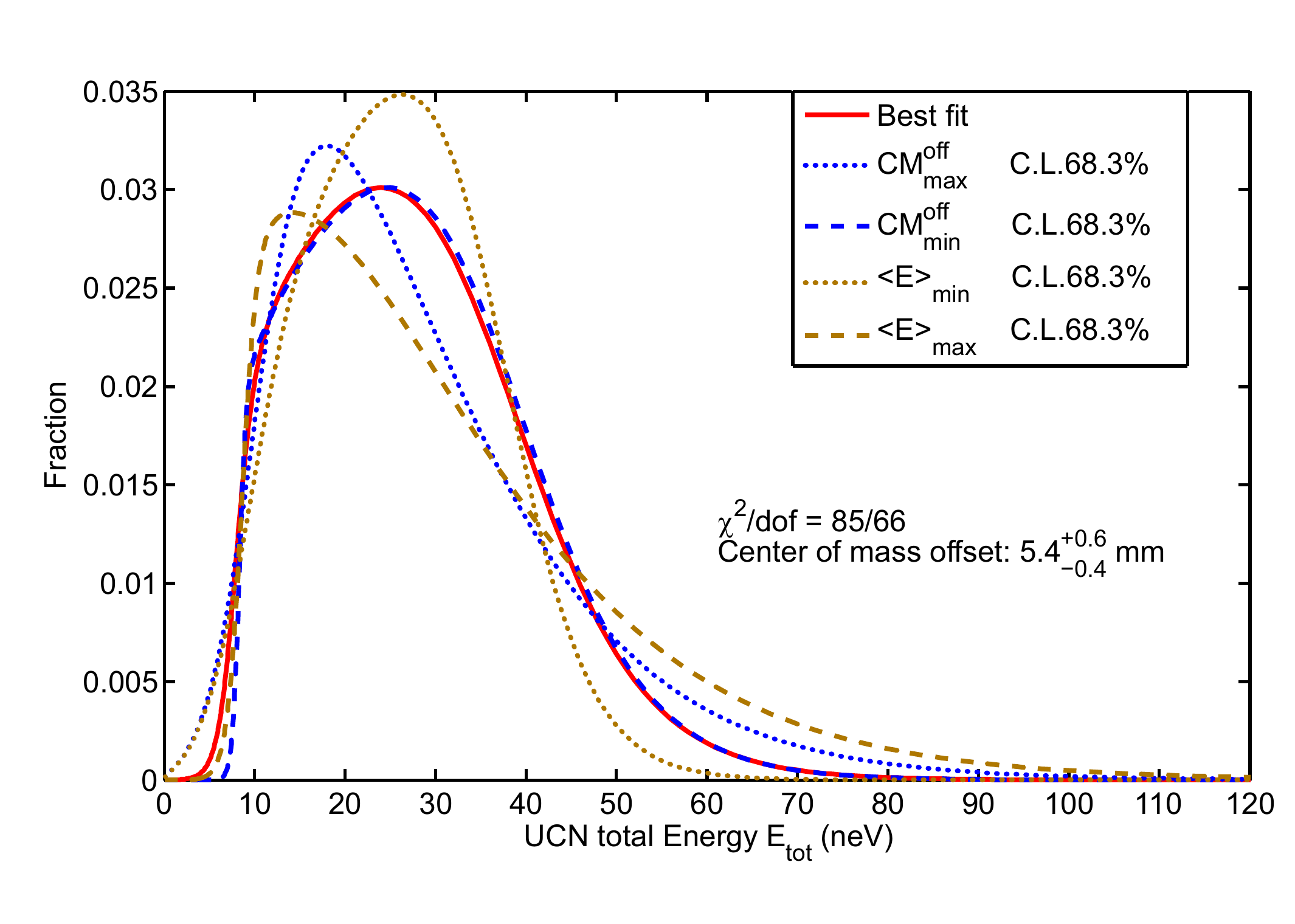}%
\caption{Normalized energy spectra of stored UCNs after \unit[220]{s} storage. The red line displays the energy spectrum obtained from the best global fit to equation\,\eqref{eq:Polarization2}, using \eqref{eq:spectrum} and \eqref{eq:phase2ndOrder}, to all three data sets, compare Fig.\,\ref{fig:UCNSE_exp}. The additional spectra shown are those having the highest and lowest mean energy (brown) and the highest and lowest center-of-mass offset (blue) within a $68.3\%$ confidence interval.}%
\label{fig:UCNSESpectrum}%
\end{figure}

A full phase coherent $\pi/2$ -- $\pi$ -- $\pi/2$ spin-flip sequence was applied to the neutrons in the cell
before the
UCN shutter was then opened again, and the UCN were detected in a pair of spin-state-specific detector channels. 
A measurement consisted of several cycles with $t_1$ increasing in steps of \unit[9]{s}. 
Figure\,\ref{fig:UCNSE_exp} shows the three measured spin-echo resonances. 
As for the simulated cases, equation\,\eqref{eq:Polarization2} was fitted to the data, using the same fitting procedure and in addition randomly varying the initial start values.
The frequency $\omega_\text{r}$ from equation\,\eqref{eqn:frequencyRot} was corrected for the difference $\Delta\omega = \omega_\text{rf} -|\gamma_\mathrm{n}|/\gamma_\mathrm{Hg}\omega_\text{Hg}$ and was also extended by a term for the second-order gradient in $z$, which was available from the field measurement: 
\begin{align}
			\omega_\text{r}(\epsilon) =&\gamma_\text{n}\left[-\left(\Part{B_z}{z}+G_z\right)\cdot \left( H/2-  \left\langle h(\epsilon)\right\rangle \right) +\right.\nonumber \\
			&\left.\frac{\partial^2 B_z}{\partial z^2}\!\cdot\!\left(\langle h^2\rangle-\left\langle (H/2 - h(\epsilon))^2\right\rangle  \right)\right]-\Delta\omega, \label{eq:phase2ndOrder}
\end{align}

\noindent where $\langle h^2\rangle =\int_{-6}^{+6}{z^2/H}\diff z = \unit[12]{cm^2}$ is the spectrum-independent quadratic expectation value.
The fit parameter $G_z$ was included to accommodate the above-mentioned common gradient offset.
As the observed polarization at $t_1=T/2$ was the same for all three measurements, we used an averaged common $\Gamma_2(\epsilon)$ for the fit.
Figure\,\ref{fig:UCNSESpectrum} shows the extracted energy spectrum.  The best fit parameters were:\linebreak $G_z\!=\!\unit[5.8(1.1)]{pT/cm}$, $a\!=\! \unit[10.0(4)]{} $, $b\!=\!\unit[1^{+0.9}_{-0.0}]{neV^{1/2}}$, $c \!=\!\unit[2.7^{+1.2}_{-0.2}]{neV^{1/2}} $, $E_0 \!=\! \unit[7.7^{+4.3}_{-1.4}]{neV} $, and $E_1\! =\! \unit[28.7^{+4.5}_{-4.5}]{neV}$, with $E_2$ fixed at $\unit[200]{neV}$. Lower bounds for $b=\unit[1]{neV^{1/2}}$ and $c=\unit[2.5]{neV^{1/2}}$ were chosen to give smooth physical edges while $\chi^2$ remained essentially constant for even lower values.
The average energy $\langle E \rangle$ of the measured spectrum is \unit[$27.6^{+4.1}_{-1.7}$]{neV}, corresponding to a mean velocity at the base of the containment vessel of $\overline{v} = \unit[230^{+16}_{-7}]{cm/s}$. The observed intrinsic depolarization time at $t_1 = T/2$ is $T_2 = \unit[506(26)]{s}$, while the calculated dephasing times, correcting for the frequency offset $\Delta \omega$ are $T_2^{\ast} = 449(8)$, $439(10)$, and \unit[279(9)]{s} for the measurements at nominal $\partial B_z/\partial z = -18$, \unit[10], \unit[38]{pT/cm} respectively. Note that $a=10.0(4)$ indicates that the intrinsic depolarization is a factor 10 higher than calculated which we attribute to higher order magnetic-field gradients or magnetic impurities of the walls of the precession cell. 

The extracted spectrum-averaged UCN to \magHg center-of-mass offset $h_\text{off} =\unit[5.4^{+0.6}_{-0.4}]{mm}$ is larger using this method than the value \unit[2.35(5)]{mm}\,\cite{Afach2014} deduced from a measurement of the ratio of the precession frequencies of \magHg{} atoms to neutrons, $R=\omega_\text{n}/\omega_\text{Hg}$, after \unit[180]{s} of storage (see Ref.\,\cite{Afach2014}).
This difference is explained by the effect on the measurement of $R$ of gravitationally enhanced depolarization, as discussed in \cite{Harris2014,Afach2015PRD}, as well as a softening of the spectrum for longer storage times.  


~\\

We have demonstrated, using simulations, that the analysis of UCN spin-echo resonance signals in combination with knowledge of the ambient magnetic field provides an excellent method by which to reconstruct the energy spectrum of a confined ensemble of neutrons. The method takes advantage of the relative dephasing of spins arising from a gravitationally induced striation of stored UCN of different energies. Based upon this, measured UCN spin-echo signals have been analyzed. They were taken in a controlled magnetic environment which allowed for long inherent coherence times and permitted a FSP time of \unit[216]{s}. From these measurements it was possible not only to extract the energy spectrum of stored UCNs, but also to determine a common gradient offset with a resolution of \unit[1.1]{pT/cm}, and thus to determine the magnetic-field gradient at this level of accuracy. We plan to improve the energy resolution  at higher energies ($50$ to \unit[200]{neV}) by repeating this measurement with a larger bottle of $H\approx\unit[40]{cm}$. 
Using an array of vector magnetometers, as described in e.g.\ Ref.\,\cite{Afach2015PRL}, will further improve this technique. Obtaining the optimal resolution will also require an improved understanding of the intrinsic-depolarization processes $\alpha(T,\epsilon)$, particularly in the limit of very low energies, i.e.\ $E\!<\!H$, including effects of different specularities of wall reflections and higher order magnetic-field gradients.  

This technique will improve the estimation of energy-spectrum dependent systematic effects in high-precision experiments such as searches for the neutron electric dipole moment\,\cite{Baker2011} and for spin-dependent forces\,\cite{Afach2014b}. Measurements for different $T$ will give access to the evolution of the spectrum $p(\epsilon,T)$ during storage, which may help to solve the neutron lifetime controversy\,\cite{Olive2014}. Alternating measurements with and without the $\pi$-pulse at $t_1=T/2$ could also provide a powerful method to improve the magnetic-field homogeneity of typical UCN spin-precession experiments, such as, for example, neutron EDM searches, where it is known that magnetic field gradients can cause false EDM signals when a co-magnetometer is used\,\cite{Pendlebury2004,Afach2015ELJD}. Using the gradient-offset extraction from a spin-echo measurement in combination with a time-resolved gradient measurement by the CsM array may allow one to correct directly for these false-EDM signals in the future.

\section*{Acknowledgments}
We would like to thank the PSI staff, in particular F.~Burri and M.~Meier, for their outstanding support.  We also gratefully acknowledge the important work carried out by the workshops throughout the collaborating institutes and the computational power provided by the PL-Grid infrastructure. One of us~(EW) benefited from a PhD fellowship of the research foundation Flanders (FWO). This research was financed in part by the Fund for Scientific Research, Flanders; grant GOA/2010/10 of KU~Leuven; the Swiss National Science Foundation Projects 200020-144473 (PSI), 200021-126562 (PSI), 200020-149211 (ETH) and 200020-140421 (Fribourg); and grants ST/K001329/1, ST/M003426/1 and ST/L006472/1 from the UK's Science and Technology Facilities Council (STFC). The original apparatus was funded by grants from the UK's PPARC (now STFC). The LPC Caen and the LPSC acknowledge the support of the French Agence Nationale de la Recherche (ANR) under reference ANR-09-BLAN-0046. Our Polish partners wish to acknowledge support from the National Science Centre, Poland, under grant no.\ UMO-2012/04/M/ST2/00556.


\input{UCNSEBibLett}
\end{document}

%% file: CommandsAndShortcuts.tex
\newlength{\bildtitel}
\newcommand{\diff}[1]{\operatorname{d}\ifthenelse{\equal{#1}{}}{\,}{\!#1}}

\newcommand{\Part}[2]{\displaystyle\frac{\partial #1}{\partial #2}}

\newcommand{\muT}{\mbox{\micro T}}

\newcommand{\mus}{\ensuremath{\micro \mathrm{s} }}

\newcommand{\magHg}{\texorpdfstring{\ensuremath{{}^{199}\text{Hg}}}{mercury-199}\xspace}

